\documentclass[12pt]{iopart}
\usepackage{iopams}
\usepackage{verbatim}
\usepackage[dvips]{epsfig}

\def\##1{{\underline #1}}
\def\~#1{{\underline {\mathcal#1}}}
\def\+#1{{{\mathcal #1}}}
\def\=#1{\underline{\underline #1}}

\def\.{\mbox{ \tiny{$^\bullet$} }}

\def\eps{\epsilon}
\def\epso{\epsilon_o}
\def\epsa{\epsilon_a}
\def\epsb{\epsilon_b}
\def\epsc{\epsilon_c}
\def\epsd{\epsilon_d}

\def\ux{\#u_x}
\def\uy{\#u_y}
\def\uz{\#u_z}

\def\muo{\mu_o}
\def\lambdao{\lambda_o}
\def\etao{\eta_o}
\def\ko{k_o}

\def\vbar{\overline{v}}

\def\cpg{\cos\gamma}
\def\spg{\sin\gamma}
\def\cpgsq{\cos^2\gamma}
\def\spgsq{\sin^2\gamma}

\def\le{\left(}
\def\ri{\right)}
\def\les{\left[}
\def\ris{\right]}

\def\ric{\right\}}

\def\chivt{\tilde\chi_v}
\def\vbar{\overline{v}}
\def\dgamma{\Delta\gamma}
\def\decayone{{\rm Im}[\alpha_1]/k_o}
\def\decaytwo{{\rm Im}[\alpha_2]/k_o}

\begin{document}

\title[Dyakonov--Tamm waves at nematic
thin film/isotropic dielectric interface]{Theory of Dyakonov--Tamm waves at the planar interface of a sculptured nematic
thin film and an isotropic dielectric material}

\author{Kartiek Agarwal,$^{1,2}$ John A. Polo, Jr.,$^{3}$ and Akhlesh Lakhtakia$^{2,4}$\footnote{Corresponding author; e-mail:akhlesh@psu.edu}}
\address{$^{1}$ Department of Electrical Engineering, Indian Institute of Technology Kanpur, Kanpur 208016, India}
\address{$^{2}$ NanoMM---Nanoenginered Metamaterials Group, Department of Engineering Science and Mechanics, Pennsylvania State University, University Park, PA 16802, USA}
\address{$^{3}$ Department of Physics and Technology, Edinboro University of Pennsylvania,
Edinboro, PA 16444, USA}
\address{$^{4}$ Department of Physics, Indian Institute of Technology Kanpur, Kanpur 208016, India}

\begin{abstract} In order to ascertain conditions for surface--wave propagation guided by the planar
interface of an isotropic dielectric material and a sculptured nematic thin film (SNTF) with 
periodic nonhomogeneity, we formulated a boundary--value problem, obtained a dispersion
equation therefrom, and numerically solved it. The surface waves obtained are Dyakonov--Tamm waves.
The angular domain formed by the directions of propagation of the Dyakonov--Tamm
waves can be very wide (even as wide as to allow propagation in every direction in the interface plane), because of the
periodic nonhomogeneity of the SNTF. A search for Dyakonov--Tamm waves is, at the present time, the 
most promising route to take for experimental veri\-f\-i\-cation of surface--wave propagation 
guided by the interface of two dielectric materials, at least one of which is  anisotropic. That would also
assist in realizing the potential of such surface waves for optical sensing of various types
of analytes infiltrating one or both of the two dielectric materials.

\vspace{0.5cm}

\end{abstract}

\maketitle

\section{Introduction}\label{intro}

In 1988, Dyakonov \cite{Dyakonov1988} theoretically predicted the propagation of electromagnetic waves
guided by the planar interface of two homogeneous dielectric materials, one of which is isotropic and
the other uniaxial with its optic axis aligned
parallel to the interface. Since then, the existence of Dyakonov surface waves has been theoretically
proved for many sets of dissimilar dielectric partnering materials, at least one of which is anisotropic
\cite{Takayama, Crasovan}. The possibility of the
anisotropic partnering material being artificially engineered, either as a photonic crystal with a short period in comparison to the wavelength \cite{Artigas2005} or as
a columnar thin film (CTF) \cite{PNL2007}, has also emerged. Just like that of many surface phenomena
\cite{AZL1}, the
significance of Dyakonov surface waves for optical sensing applications is obvious: the disturbance of the
constitutive properties of one or both of the two partnering materials---due to, say, infiltration by
any analyte---could measurably change the characteristics
of the chosen Dyakonov surface wave.

However, the directions of propagation of Dyakonov surface waves parallel to any suitable planar
interface are confined to a very narrow angular domain, typically of the order of a degree or
less \cite{Takayama,NPL-motl}. Not surprisingly, Dyakonov surface waves still remain  to be experimentally observed.
Clearly then, the potential of Dyakonov surface waves cannot be realized if they
cannot be demonstrably excited; a wider angular domain is needed.

Recently, Lakhtakia and Polo \cite{LP-jeosrp} examined surface--wave propagation  guided by
the planar interface of an isotropic dielectric material and a chiral sculptured thin film \cite{LMbook} with
its direction of periodic nonhomogeneity normal to the planar interface. The authors
named the surfaces waves  after Dyakonov  and  Tamm, the latter being the
person who indicated the possibility of finding surface states
of electrons at exposed planes of crystals and other periodic materials \cite{Tamm,Ohno1990}. The angular domain
formed by the directions of propagation of the Dyakonov--Tamm waves turned out to be very wide,
as much as $98^\circ$ in one case for which realistic constitutive parameters of the two
partnering materials were employed \cite{LP-jeosrp}. This implies that
 Dyakonov--Tamm waves could be detected much more easily than Dyakonov surface waves.

 Chiral sculptured thin films (STFs) are structurally chiral materials:\footnote{An object is said to be chiral 
 if it cannot be made to coincide with its mirror-image by translations
and/or rotations.
There are two types of chiral materials: (i) microscopically or molecularly chiral materials, and (ii)
structurally chiral materials. The first type of 
chiral materials either have chiral molecules or are composite materials made by embedding, e.g., electrically small helixes 
in a host material. Materials comprising chiral molecules have been known for about two hundred years, as a perusal
of an anthology of milestone papers \cite{NOA} will show to the interested reader. These materials are generally
(optically) isotropic \cite{Charney}. Composite materials
comprising electrically small 
\cite{vdH,TGM} chiral inclusions were first reported in 1898 \cite{Bose}, and these materials can be either isotropic
\cite{Ro,Bel,Alvaro} or anisotropic \cite{Whites}. The first type of chiral materials can be considered as either homogeneous or nonhomogeneous continuums
at sufficiently low frequencies.
 In contrast,
the second type of chiral materials can only be nonhomogeneous and anisotropic continuums at the length--scales
of interest, their constitutive parameters varying periodically in a chiral manner about a fixed axis. Take away the nonhomogeneity of a 
structurally chiral material, and its (macroscopic) chirality will also vanish \cite{LMbook,Chan}.}
  the permittivity dyadic rotates at a uniform
 rate along the direction of nonhomogeneity. As a result, chiral STFs are periodically nonhomogeneous.
 Thus, in a chiral STF, structural chirality and periodic nonhomogeneity are inseparable.
 Is the huge angular existence domain of Dyakonov--Tamm waves for the case
 investigated by Lakhtakia and Polo \cite{LP-jeosrp} due to  the periodicity
 or due to the structural chirality of the chiral STF? In order to answer this question, we devised
 a surface--wave--propagation problem wherein the chiral STF is replaced by
 a periodically nonhomogeneous and anisotropic material that is not structurally chiral. Specifically,
 we replaced the chiral STF with a sculptured nematic thin film (SNTF) \cite{LMbook}.

\begin{figure}
    \begin{center}
    \begin{tabular}{c}
    \includegraphics[height=5cm]{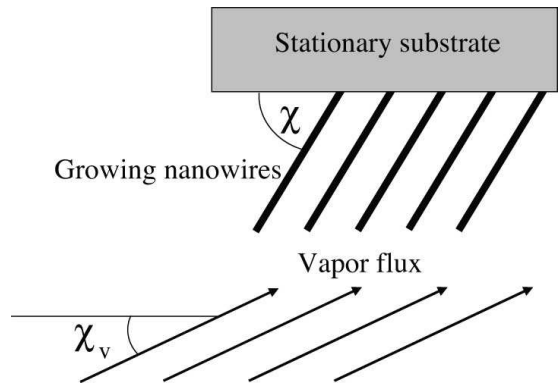}
    \end{tabular}
    \end{center}
    \caption{ \label{CTFgrowth} Schematic of a collimated vapor flux responsible for the growth
    of tilted straight nanowires growing at an angle $\chi\geq\chi_v$.}
 \end{figure}

\begin{figure}
    \begin{center}
    \begin{tabular}{c}
    \includegraphics[height=6.3cm]{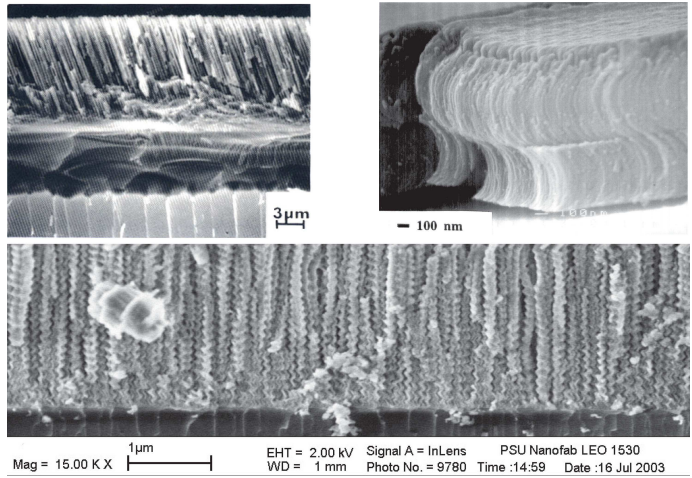}
    \end{tabular}
    \end{center}
    \caption{ \label{STFs} Scanning-electron-microscope images of sculptured thin films. Top left: columnar thin film; top right: sculptured
    nematic thin film; bottom: chiral sculptured thin film. Courtesy: Russell Messier and Mark Horn. }
 \end{figure}

 CTFs, chiral STFs, and SNTFs are all fabricated by physical vapor deposition \cite{LMbook,
 HWbook,Mattox}. In the simplest
 implementation of this technique, a boat containing a certain material, say titanium oxide, is
 placed in an evacuated chamber. Under appropriate conditions, the material evaporates towards
 a substrate such that the vapor flux is highly collimated. The collimated vapor
 flux coalesces on the substrate as an ensemble
 of more or less identical and parallel nanowires,
 due to self--shadowing.   If the substrate is held stationary during deposition,
 a CTF grows wherein the nanowires are straight
 and aligned at an angle $\chi\geq\chi_v$ with respect to the substrate plane, where $\chi_v$ is the
angle between the collimated vapor flux and the substrate plane, as shown in Fig.~\ref{CTFgrowth}. If the substrate is rotated about an axis
passing normally through it, a chiral STF comprising helical nanowires grows. Finally, if the
substrate is rocked about an axis tangential to the substrate plane, $\chi_v$ and $\chi$ are not constant
and an SNTF grows. Scanning-electron-microscope images of a CTF, an SNTF, and a chiral STF are presented
in Fig.~\ref{STFs}.

This paper is organized as follows. Section~\ref{form} presents the boundary--value problem and the dispersion equation for the propagation of 
Dyakonov--Tamm waves guided
by the interface of a homogenous, isotropic dielectric
 material and a periodically nonhomogeneous SNTF. Section~\ref{res} contains numerical results when the SNTF is chosen to be made of titanium oxide \cite{PNL2007,HWH1998}. An $\exp(-i\omega t)$ time--dependence is implicit, with $\omega$ denoting the angular frequency. The free--space wavenumber, the free--space wavelength, and the intrinsic impedance of free space are denoted by $\ko=\omega\sqrt{\epso\muo}$, $\lambdao=2\pi/\ko$, and $\etao=\sqrt{\muo/\epso}$, respectively, with $\muo$ and $\epso$ being  the permeability and permittivity of free space. Vectors are underlined, dyadics   underlined twice; column vectors are underlined and enclosed within square brackets, while matrixes are underlined twice and similarly bracketed. Cartesian unit vectors are identified as $\ux$, $\uy$ and $\uz$. The dyadics employed in the following sections can be treated as 3$\times$3 matrixes \cite{Chen}.

\section{Formulation}\label{form}
\subsection{Geometry and permittivity}

Let the half--space $z\leq0$ be occupied by an isotropic, homogeneous, nondissipative, dielectric
material of refractive index $n_s$. The region $z\geq0$ is occupied by an SNTF with a
periodically nonhomogeneous  permittivity dyadic given by \cite{LMbook}
\begin{equation}
\=\epsilon(z)= \epso \, \=S_z(\gamma)\cdot\=S_y(z)\cdot\=\epsilon^\circ_{
ref}(z)\cdot\=S_y^T(z)\cdot\=S_z^T(\gamma)
\, , \quad z\geq 0 \,.
\end{equation}
The
dyadic function
\begin{eqnarray}
\nonumber
\=S_z(\gamma)&=& \le \ux\ux + \uy\uy \ri \cos\gamma \\
&&+\, \le \uy\ux - \ux\uy \ri\sin\gamma+\uz\uz \,\label{SzDyadic}
\end{eqnarray}
contains  $\gamma$ as an angular offset, and the superscript $^T$ denotes the transpose.
The dyadics
\begin{equation}
\=S_{y}(z)=(\ux\ux+\uz\uz) \cos\left[\chi (z)\right] +
(\uz\ux - \ux\uz) \sin \left[\chi (z)\right] +\uy\uy
\end{equation}
and
\begin{equation}
\=\epsilon^{\circ} _{ref}(z)=\epsa(z)\,\uz\uz+\epsb(z)\,\ux\ux+\epsc(z)\,\uy\uy
\end{equation}
depend on the vapor incidence angle
\begin{equation}
\chi_v(z) = {\tilde\chi}_v + \delta_v\,\sin\left(\frac{\pi z}{\Omega}\right)\,
\end{equation}
that varies sinusoidally as a function of $z$ with period $2\Omega$.
Whereas ${\tilde\chi}_v\geq0$, we have to ensure that $\chi_v(z)\in\left(0,\pi/2\right]$.

As no experimental data connecting $\eps_{a,b,c}(z)$ and $\chi(z)$
to $\chi_v(z)$ have been reported for SNTFs ($\delta_v\ne 0$), we decided to use
available data for CTFs ($\delta_v=0$).
Optical
characterization experiments on CTFs of titanium oxide
 at $\lambdao=633$~nm \cite{HWH1998} lead to the following expressions  for
 our present purpose:
\begin{equation}
\label{Hodg1}
\left. \begin{array}{ll}
\eps_a(z)  = \left[ 1.0443 + 2.7394\,v(z) -1.3697\,v^2(z)\right]^2\\[5pt]
\eps_b(z) = \left[ 1.6765 + 1.5649\,v(z) -0.7825\,v^2(z)\right]^2\\[5pt]
\eps_c(z)  = \left[ 1.3586 +  2.1109\,v(z) -1.0554\,v^2(z)\right]^2\\[5pt]
\chi (z) = \tan^{-1}\left[2.8818\,\tan\chi_v(z)\right]\\[5pt]
v(z) =2\chi_v(z)/\pi
\end{array}\ric\,.
\end{equation}
Let us note that
the foregoing expressions---with $v(z)$ independent of $z$---are applicable to CTFs
produced by one particular experimental apparatus, but may have to be modified for  CTFs produced by other researchers on
different apparatuses.

Without loss of generality, we take the Dyakonov--Tamm wave to propagate parallel to the $x$ axis in the plane $z=0$.
There is no dependence on the $y$ coordinate, whereas the Dyakonov--Tamm wave must
attenuate as $z\to\pm\infty$.

\subsection{Field representations}
In the region $z \leq 0$, the wave vector may be written as
\begin{equation}
\#k_s=\kappa\, \ux -\alpha_s\,\uz\,,
\end{equation}
where
\begin{equation}
\kappa^2+\alpha_s^2=\ko^2\, n_s^2\,,
\end{equation}
 $\kappa$ is positive and real--valued for unattenuated
propagation along the $x$ axis, and
 ${\rm Im}\left[\alpha_s\right]>0$ for attenuation as $z\to-\infty$.
Accordingly, the field phasors in the region $z\leq 0$ may be written as \cite{LP-jeosrp}
\begin{equation}
\#E(\#r)= \left[A_{s}\,\uy + A_{p}\left( \frac{\alpha_s}{\ko}\,\ux +\frac{\kappa}{\ko}\, \uz\right)\right]
\exp(i\#k_s\cdot\#r)\,,\quad z \leq 0\,,
\label{eqn:Esub}
\end{equation}
and
\begin{equation}
\#H(\#r)=\etao^{-1}\left[ A_{s} \left(\frac{\alpha_s}{\ko}\,\ux+\frac{\kappa}{\ko}\,\uz\right) -
A_{p}\,n_s^2\,\uy\right] \exp(i\#k_s\cdot\#r)\,,\quad z \leq 0\,,\label{eqn:Hsub}
\end{equation}
where $A_{s}$ and $A_{p}$ are unknown scalars representing the amplitudes of $s$-- and
$p$--polarized components.

 The field representation in the region $z\geq0$ is more complicated. It is appropriate to
write
\begin{equation}
\left.\begin{array}{l}
\#E(\#r)=\#e(z)\,\exp(i\kappa x)\\
\#H(\#r)=\#h(z)\,\exp(i\kappa x)
\end{array}\right\}\,,\qquad
z\geq0\,,
\end{equation}
and create the column vector
\begin{equation}
\left[\#f(z)\right]= \left[e_x(z)\quad e_y(z) \quad h_x(z)\quad h_y(z)\right]^T\,.
\end{equation}
This column vector satisfies the matrix differential equation \cite{LMbook}
\begin{equation}
\label{MODE}
\frac{d}{dz}\left[\#f(z)\right]=i \left[\=P(z,\kappa)\right]\cdot\left[\#f(z)\right]\,,
\quad z>0\,,
\end{equation}
where the 4$\times$4 matrix
 \begin{eqnarray}
\nonumber
&&
[\=P(z,\kappa)]=
\end{eqnarray}
\begin{eqnarray}
\nonumber
\omega\,\les\begin{array}{cccc}
0 & 0 & 0 & \muo \\[4pt]
0 & 0 & -\muo & 0 \\[4pt]
 \epso\left[\epsc(z)-\epsd(z)\right]\cpg\,\spg & -\epso\left[\epsc(z)\cpgsq
+\epsd(z)\spgsq\right] & 0 & 0\\[4pt]
\epso\left[\epsc(z)\spgsq +\epsd(z)\cpgsq\right] & -\epso\left[\epsc(z)-
\epsd(z)\right]\cpg\,\spg & 0 & 0
\end{array}\ris
\end{eqnarray}
\begin{eqnarray}
\nonumber
&& \qquad
+\,\kappa\,\frac{\epsd(z)\,\left[\epsa(z)-\epsb(z)\right]}{\epsa(z)\,\epsb(z)}\,
\sin\chi(z)\cos\chi(z)\,
\les\begin{array}{cccc}
\cpg  & \spg  & 0 & 0\\[4pt]
0 & 0 & 0 & 0\\[4pt]
0 & 0 & 0& -\spg \\[4pt]
0 & 0 & 0 & \cpg
\end{array}\ris
\\[5pt]
&& \qquad
+
\les\begin{array}{cccc}
0 & 0 & 0 & -\,\frac{\kappa^2}{\omega\epso}\,\frac{\epsd(z)}{\epsa(z)\,\epsb(z)}\\[4pt]
0 & 0 & 0 & 0 \\[4pt]
0 & \frac{\kappa^2}{\omega\muo}& 0 & 0\\[4pt]
0 & 0 & 0 & 0
\end{array}\ris\,
\label{Pdef}
\end{eqnarray}
and
\begin{equation}
\epsd(z)=\frac{\epsa(z)\epsb(z)}{\epsa(z)\,\cos^2\chi(z)+\epsb(z)\,\sin^2\chi(z)}\,.
\end{equation}
For $\sin\gamma=0$, (\ref{MODE}) splits into two autonomous matrix differential equations,
each employing a 2$\times$2 matrix \cite{ML2008}.

Equation (\ref{MODE}) has to be solved numerically. We used the piecewise uniform
approximation technique \cite{LMbook}
to determine the matrix $[\=N]$ which appears in the relation
\begin{equation}
[\#f(2\Omega)]=[\=N]\cdot[\#f(0+)]
\end{equation}
to characterize the optical response of one period of the chosen SNTF. By virtue of the Floquet--Lyapunov theorem
\cite{YS75}, we can define a matrix $[\=Q]$ such that
\begin{equation}
[\=N] = \exp\left\{i2\Omega[\=Q]\right\}\,.
\end{equation}
Both $[\=N]$ and $[\=Q]$ share the same eigenvectors, and their eigenvalues
are also related. Let $[\#t]^{(n)}$, $(n=1,2,3,4)$, be the eigenvector corresponding
to the  $n$th eigenvalue $\sigma_n$ of $[\=N]$; then, the corresponding eigenvalue
$\alpha_n$ of $[\=Q]$
is given by
\begin{equation}
\alpha_n = -i\frac{\ln \sigma_n}{2\Omega}\,.
\end{equation}

\subsection{Dispersion equation for Dyakonov--Tamm wave}
For the Dyakonov--Tamm wave to propagate along the $x$ axis, we must ensure that ${\rm Im}[{\alpha_{1,2}}]>0$, and set
\begin{equation}
[\#f(0+)]= \left[\,[\#t]^{(1)}\quad [\#t]^{(2)}\,\right]\cdot
\left[\begin{array}{c} B_1\\ B_2\end{array}\right]\,,
\end{equation}
where $B_1$ and $B_2$ are unknown scalars;
the other two eigenvalues of $[\=Q]$ describe waves that amplify as $z\to\infty$
and cannot therefore contribute to the Dyakonov--Tamm wave.
At the same time,
\begin{equation}
[\#f(0-)]=\left[\begin{array}{cc}
0 &\frac{\alpha_s}{\ko}\\[5pt]
1&0\\[5pt]
\frac{\alpha_s}{\ko}\,\etao^{-1}&0\\[5pt]
0&-n_s^2\,\etao^{-1}
\end{array}\right]\cdot
\left[\begin{array}{c} A_s\\ A_p\end{array}\right]\,,
\end{equation}
by virtue of (\ref{eqn:Esub}) and (\ref{eqn:Hsub}). Continuity of the tangential
components of the electric and magnetic field phasors across the plane
$z=0$ requires that
\begin{equation}
[\#f(0-)]=[\#f(0+)]\,,
\end{equation}
which may be rearranged as
\begin{equation}
[\=M]\cdot\left[\begin{array}{c}A_s\\A_p\\B_1\\B_2\end{array}\right]=
\left[\begin{array}{c}0\\0\\0\\0\end{array}\right]\,.
\end{equation}
For a nontrivial solution, the 4$\times$4 matrix $[\=M]$ must be singular,
so that
\begin{equation}
{\rm det}\,[\=M]= 0
\label{eq:DTdisp}
\end{equation}
is the dispersion equation for the Dyakonov--Tamm wave. The value of $\kappa$ satisfying
(\ref{eq:DTdisp}) was obtained by employing the Newton--Raphson method \cite{Jaluria}.

\section{Numerical Results and Discussion}\label{res}

We set  $\lambdao=633$ nm in accordance with (\ref{Hodg1}), since calculations were performed only for CTFs or SNTFs composed of titanium oxide. All calculations for 
the periodically nonhomogeneous SNTFs were performed for $\chivt=19.1^\circ$ and $\Omega=197$ nm with two oscillation amplitudes: $\delta_v=7.2^\circ$ and $16.2^\circ$.  For comparison, calculations were performed on CTFs for $\chi_v(z)\equiv7.2^\circ$ and $19.1^\circ\,\forall z\geq 0$.  The former is the approximate lower limit of $\chi_v$ obtainable with current STF technology.   For a chosen set  of values of $n_s$, $\tilde\chi_v$, and $\delta_v$, 
surface--wave propagation was found to occur in four separate $\gamma$-ranges:  $\gamma\in[\pm\gamma_m-\dgamma/2,\pm\gamma_m+\dgamma/2]$ and $\gamma\in[\pm\gamma_m+180^\circ-\dgamma/2,\pm\gamma_m+180^\circ+\dgamma/2]$.  Here the angle $\gamma_m\in(0^\circ,90^\circ)$ is used to describe the mid--point of 
a $\gamma$-range while $\dgamma\leq 90^\circ$ is the extent of that range.  Since the surface--wave 
characteristics are the same in all four $\gamma$-ranges, results are displayed only for $\gamma\in[\gamma_m-\Delta\gamma/2,\gamma_m+\Delta\gamma/2]$.

The magnitude of the phase velocity of the Dyakonov--Tamm wave was compared with that of the phase
velocity of the bulk wave in the isotropic partnering material.  For this purpose, we defined the
relative phase speed
\begin{equation}
\overline{v}= v_{DT}\,n_s\sqrt{\epso\muo}\,,
\end{equation}
where $v_{DT}=\omega/\kappa $ is the phase speed of the Dyakonov--Tamm wave.

\begin{figure}
    \begin{center}
    \begin{tabular}{c}
    \includegraphics[height=7.3cm]{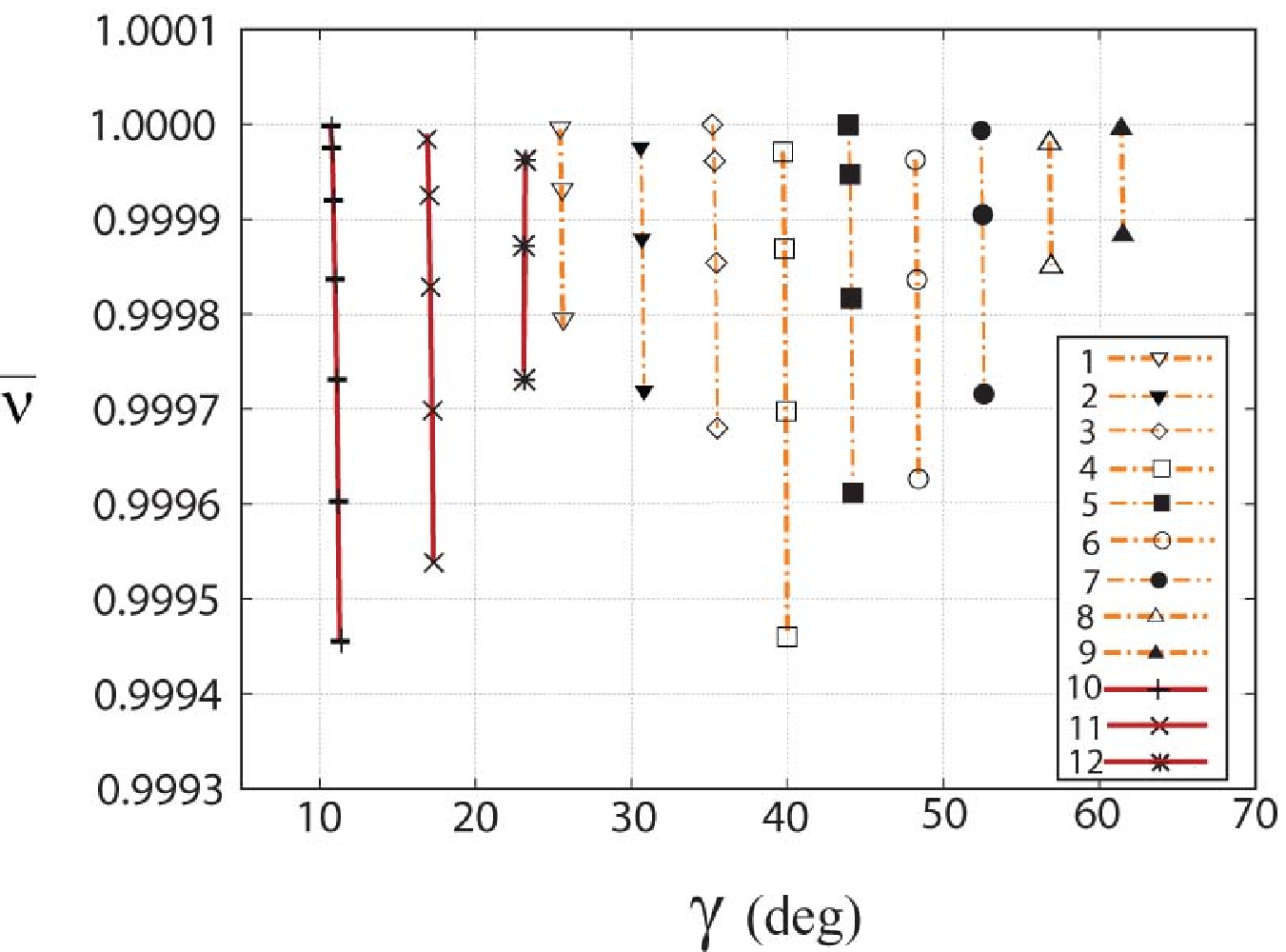}
    \end{tabular}
    \end{center}
    \caption{ \label{Fig:v_CTFs} Relative phase speed $\overline{v}$ as a function
    of $\gamma$ when $\delta_v = 0^\circ$. The values of ${\tilde\chi}_v$ are as follows: (1)-(9) $7.2^\circ$, (10)-(12) $19.1^\circ$. The values of $n_s$ are as follows: (1) 1.57, (2) 1.59, (3) 1.61, (4) 1.63, (5) 1.65, (6) 1.67, (7) 1.69, (8) 1.71, (9) 1.73, (10) 1.80, (11) 1.82, (12) 1.84. }
 \end{figure}

Let us begin the presentation and discussion of results with the characteristics of Dyakonov waves
guided by the interface of a CTF ($\delta_v=0^\circ$) and an isotropic dielectric material \cite{PNL2007}.
Figure \ref{Fig:v_CTFs} shows plots of $\vbar$ versus $\gamma$ with:  $\chi_v=7.2^\circ$ for $n_s=1.57$,
 1.59, 1.61, 1.63, 1.65, 1.67, 1.69, 1.71, and 1.73; and  $\chi_v=19.1^\circ$ for $n_s=1.80$, 1.82, and 1.84.  
Three  characteristics of Dyakonov waves can be garnered from this figure:
\begin{itemize}
\item[(i)] the nearly vertical curves demonstrate the extremely small width $\dgamma$
of the $\gamma$-range, leading to very narrow angular existence domains as we remarked upon in Section~\ref{intro};
\item[(ii)] higher values of $n_s$ result in higher values of $\gamma_m$; and
\item[(iii)] lower values of $\chi_v$ also result in higher values of $\gamma_m$.
\end{itemize}

\begin{figure}
    \begin{center}
    \begin{tabular}{c}
    \includegraphics[height=7.3cm]{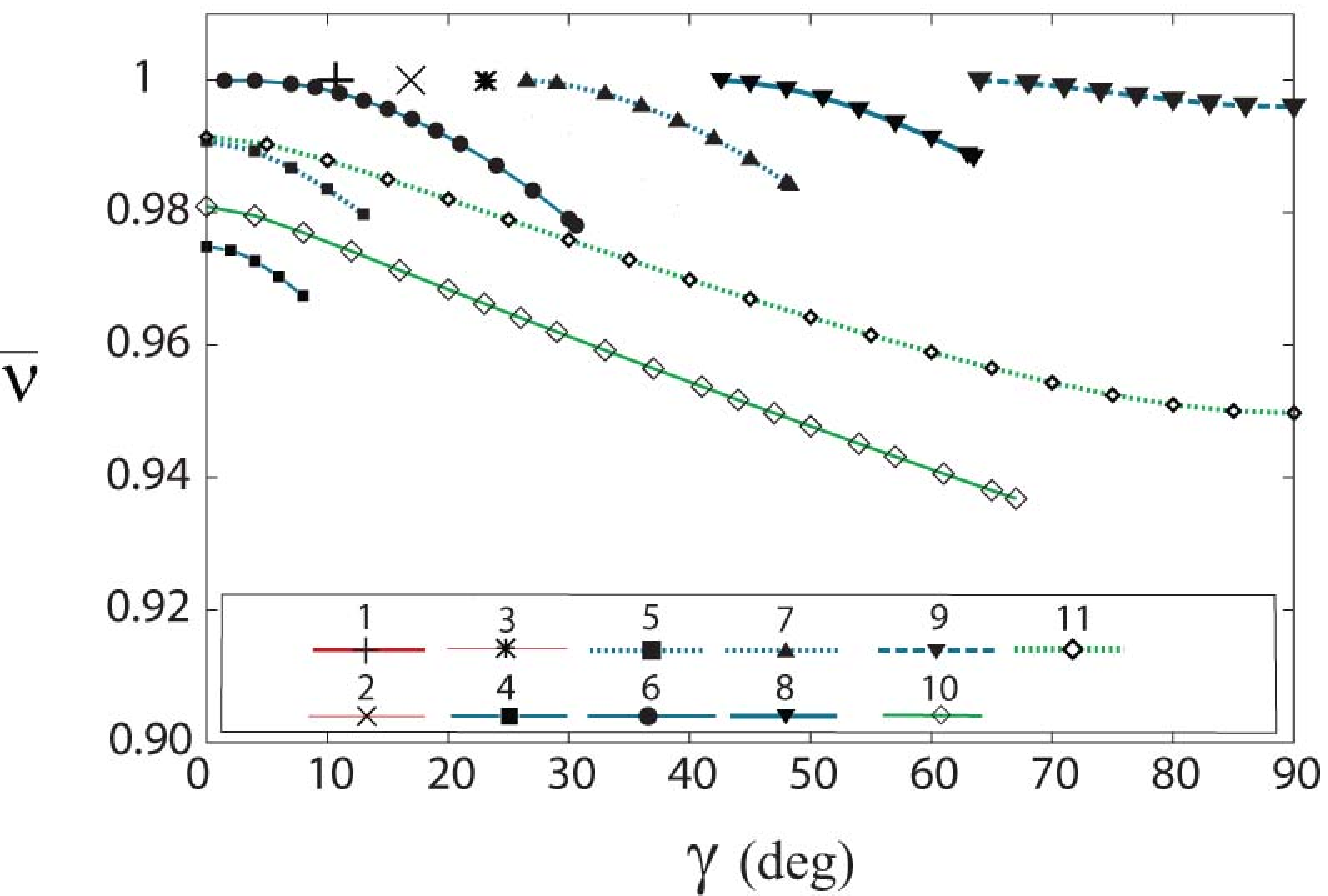}
    \end{tabular}
    \end{center}
    \caption{ \label{Fig:v_chiv_19p1} Relative phase speed $\overline{v}$ as a function
    of $\gamma$ when ${\tilde\chi}_v=19.1^\circ$. The values of
    $\delta_v$ are as follows: (1)-(3) $0^\circ$, (4)-(9)
    $7.2^\circ$,  (10) and (11) $16.2^\circ$. The values of $n_s$ are as follows:
    (1) 1.80, (2) 1.82, (3) 1.84, (4) 1.73, (5) 1.77, (6) 1.82  (7) 1.88, (8) 1.92, (9) 1.96, (10) 1.80, (11) 1.84.
    } \end{figure}

Figure \ref{Fig:v_chiv_19p1} shows $\vbar$ as a function
of $\gamma$   when ${\tilde\chi}_v=19.1^\circ$ for 
three groups of parameters:
\begin{itemize}
\item $\delta_v=0^\circ$ for $n_s= 1.80$, 1.82, and 1.84;
\item  $\delta_v=7.2^\circ$ for $n_s=1.73$, 1.77, 1.82, 1.88, 1.92, and 1.96; and 
\item $\delta_v=16.2^\circ$ for $n_s=1.80$ and $1.84$.  
\end{itemize} The first group has a CTF as the anisotropic partnering material,
whereas the second and the third groups have a periodically nonhomogeneous SNTF
serving that role.
Immediately apparent is the \emph{dramatic increase} in $\dgamma$ brought about by the introduction of sinusoidal oscillation in the vapor incidence angle.  With $\dgamma$ on the order of tens of degrees
when an SNTF is the anisotropic partnering material, the width of the $\gamma$-range is orders of magnitude greater than that obtained with a CTF as the anisotropic partnering material.  As $n_s$ increases, the   $\gamma$-range supporting 
surface--wave propagation widens and the mid-point $\gamma_m$ shifts to higher values.  When $\delta_v=16.2^\circ$ and $n_s$ is increased to 1.84, $\dgamma$ actually increases to $90^\circ$, which means that the four separate 
$\gamma$-ranges merge into one, thereby allowing surface--wave propagation to
occur for any value of $\gamma$.  

Figure \ref{Fig:v_chiv_19p1} also indicates that the average value of $\vbar$ over the $\gamma$-range
rises as $n_s$ increases, for both $\delta_v=7.2^\circ$ and $16.2^\circ$.  For a given value of $\gamma$, $\vbar$ takes on similar values for CTFs and SNTFs.

\begin{figure}
    \begin{center}
    \begin{tabular}{cc}
    \includegraphics[width=7.3cm]{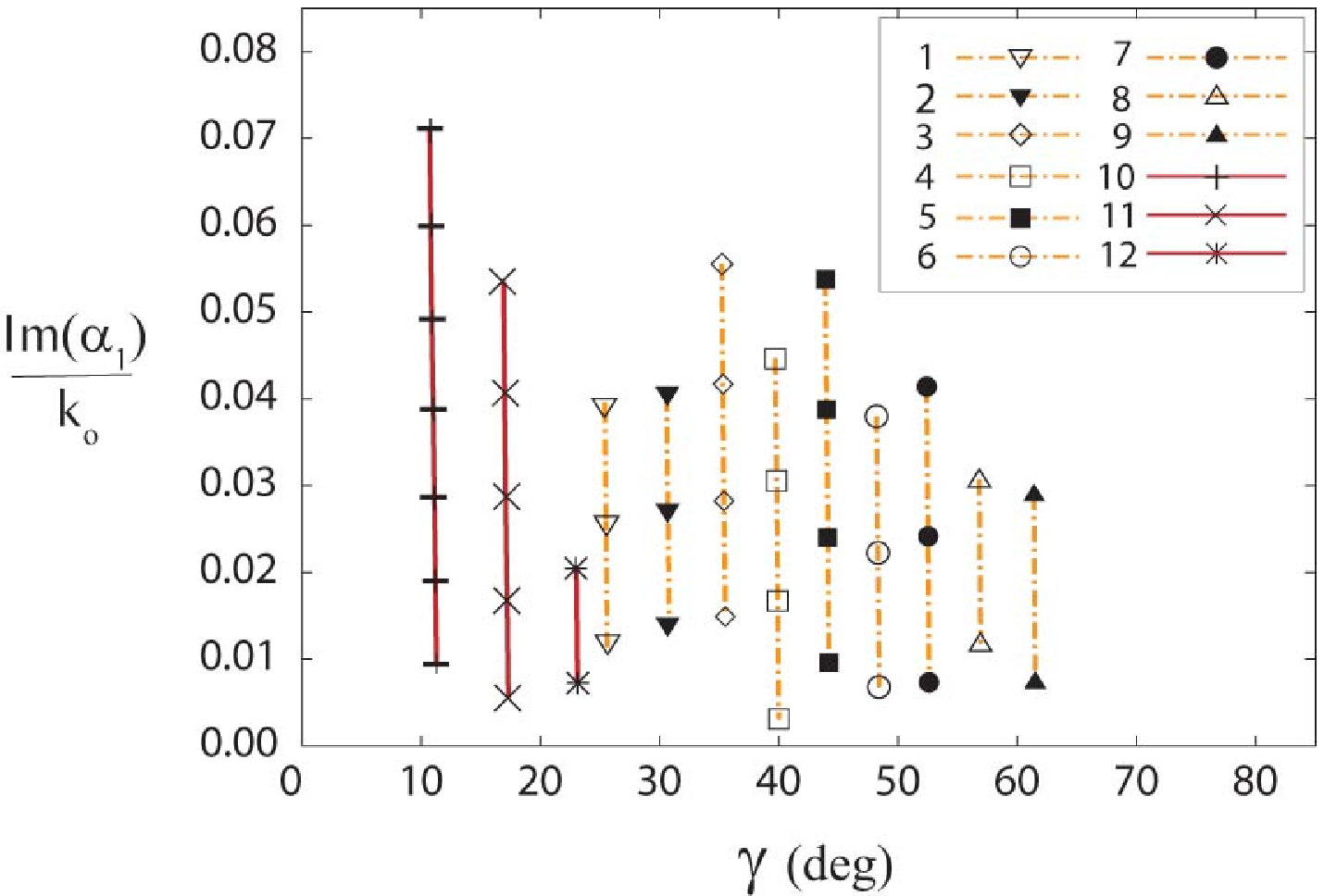}&\includegraphics[width=7.3cm]{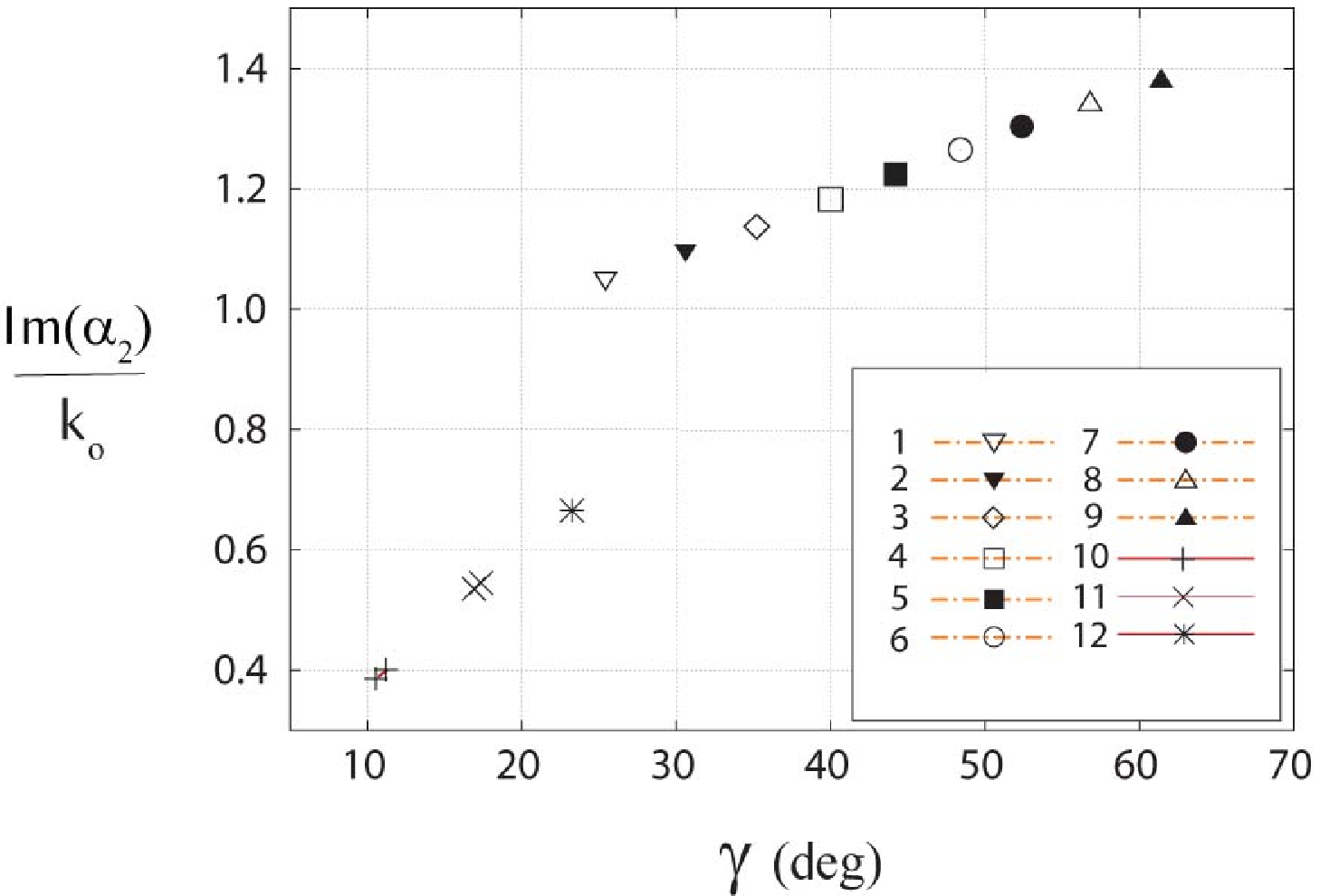}\\
    (a)&(b)
    \end{tabular}
    \end{center}
    \caption{ \label{Fig:decay_CTFs} Normalized decay constants   $\decayone$ and   $\decaytwo$ as  functions of $\gamma$ 
    when $\delta_v = 0^\circ$.  The values of ${\tilde\chi}_v$ are as follows: (1)-(9) $7.2^\circ$, (10)-(12) $19.1^\circ$. The values of $n_s$ are as follows: (1) 1.57, (2) 1.59, (3) 1.61, (4) 1.63, (5) 1.65, (6) 1.67, (7) 1.69, (8) 1.71, (9) 1.73, (10) 1.80, (11) 1.82, (12) 1.84.}
 \end{figure}

The localization of the surface wave about the bimaterial interface is described by the decay constants ${\rm Im}[\alpha_1]$ and ${\rm Im}[\alpha_2]$ in the anisotropic partnering material and by ${\rm Im}[\alpha_s]$ in the isotropic partnering
material.  Figure \ref{Fig:decay_CTFs} displays the two decay constants in the CTF normalized to the free-space wavenumber, $\decayone$ and $\decaytwo$, as  functions of $\gamma$ for $\chi_v=7.2^\circ$ and $19.1^\circ$ for the same values of $n_s$ as in Figure \ref{Fig:v_CTFs}.  The values of $\decayone$ show considerable variation over the  $\gamma$-range for a given set of $\chi_v$ and $n_s$ even though the $\gamma$-range is
very narrow, as shown by the nearly vertical curves in Figure \ref{Fig:decay_CTFs}a.  Unlike $\decayone$,   $\decaytwo$ shows little variation over the $\gamma$-range and, for most values of $n_s$, appears as a single point in Figure \ref{Fig:decay_CTFs}b.  Furthermore, $\decaytwo$ is roughly a linear function of $\gamma$ with a positive slope, for a given value of $\chi_v$.  The slope  increases as $\chi_v$ increases.

\begin{figure}
    \begin{center}
    \begin{tabular}{cc}
    \includegraphics[width=7.3cm]{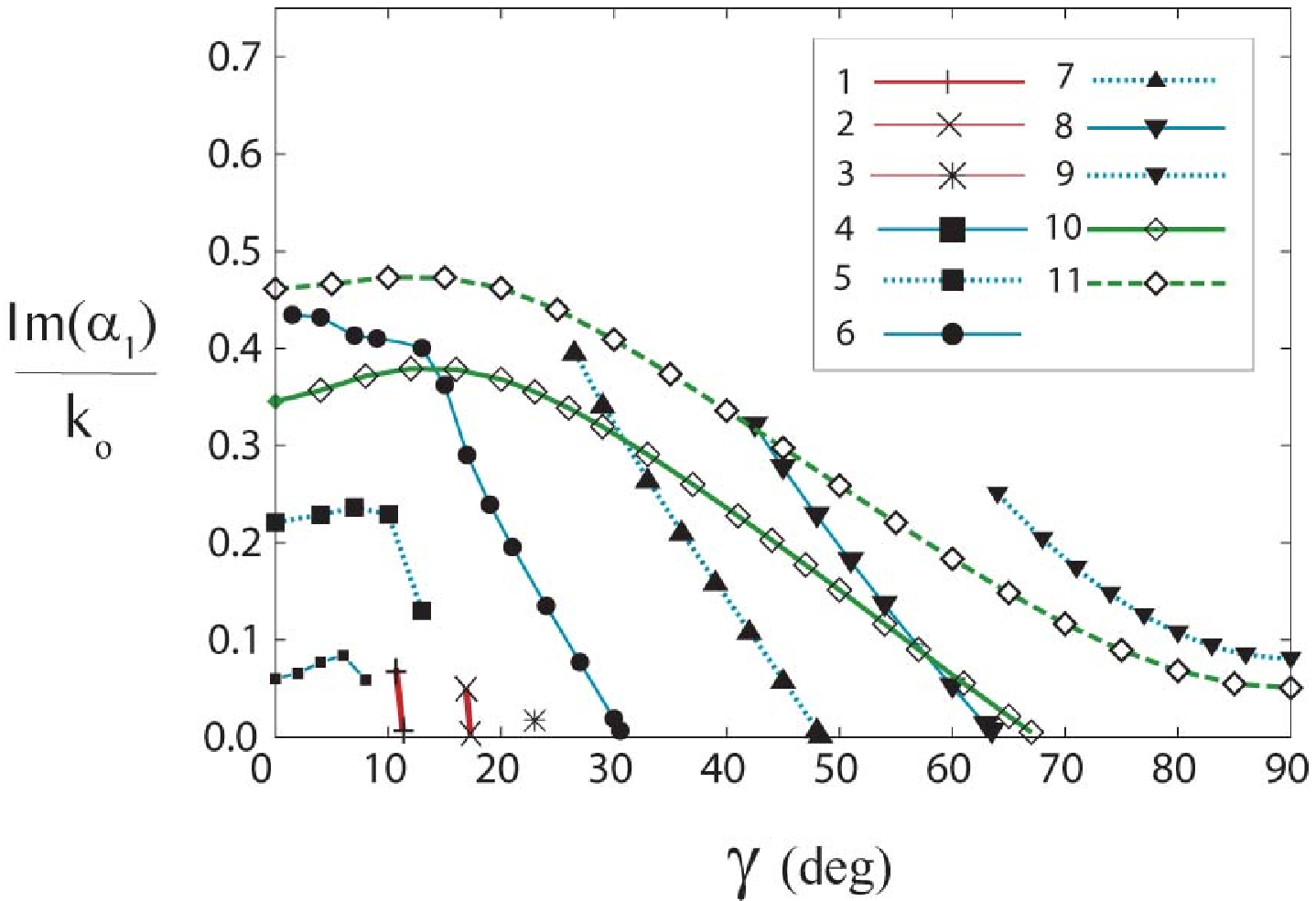}&\includegraphics[width=7.3cm]{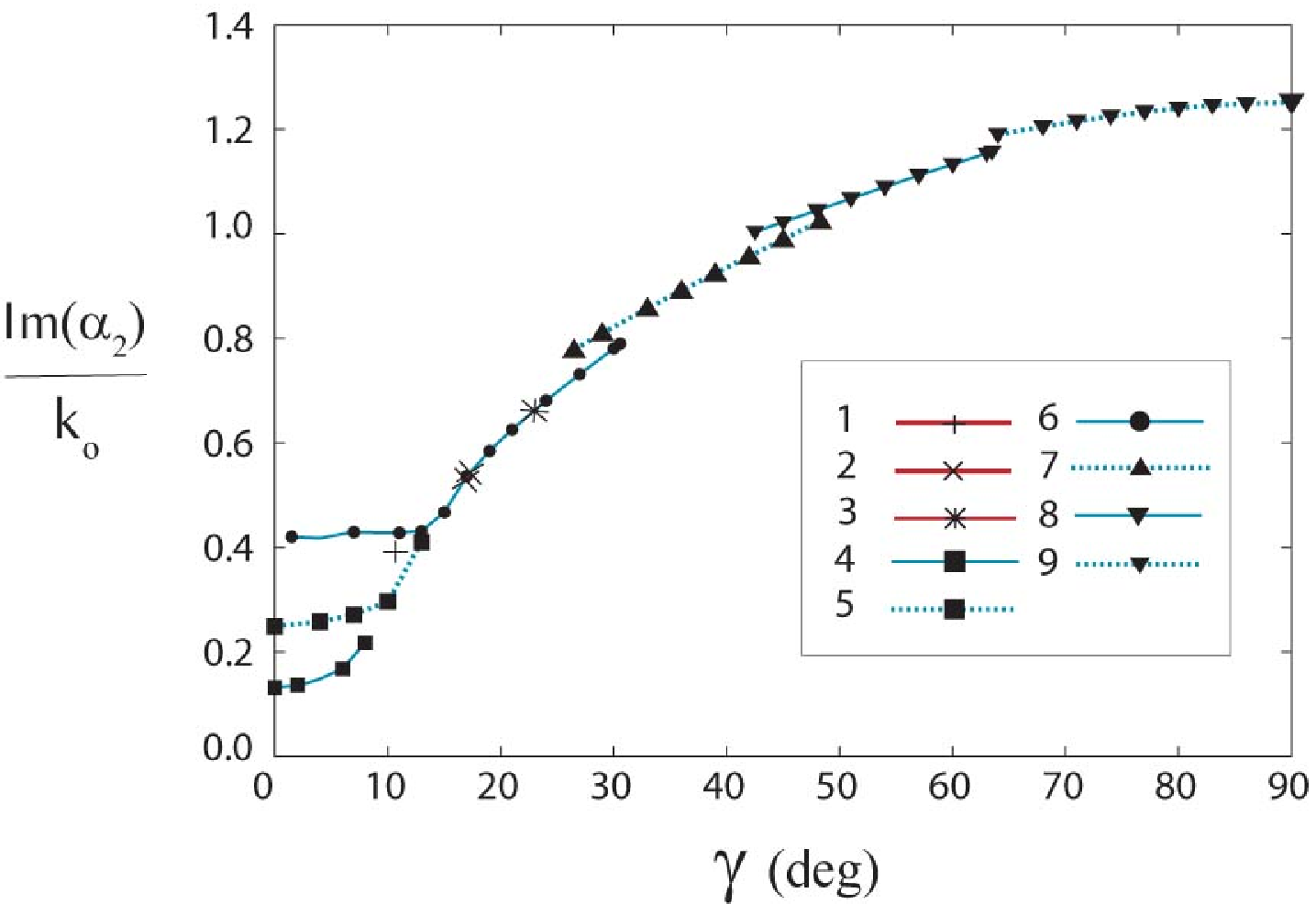}\\
    (a)&(b)
    \end{tabular}
    \end{center}
    \caption{ \label{Fig:decay_SNTF} Normalized decay constants $\decayone$ and $\decaytwo$ as functions
    of $\gamma$ when $\tilde{\chi}_v=19.1^\circ$.  The values of $\delta_v$ are as follows:  (1)-(3) $0^\circ$, (5)-(9) $7.2^\circ$, (10) and (11) $16.2^\circ$.  The values of $n_s$ are as follows:  (1) 1.80, (2) 1.82, (3) 1.84, (4) 1.73, (5) 1.77, (6) 1.82, (7) 1.88, (8) 1.92, (9) 1.96, (10) 1.80, (11) 1.84.}
 \end{figure}

\begin{figure}
    \begin{center}
    \begin{tabular}{c}
    \includegraphics[height=7.3cm]{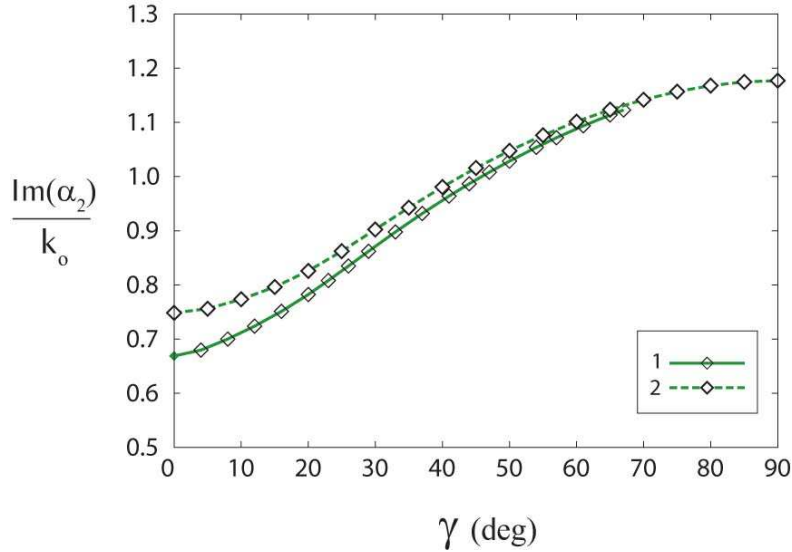}
    \end{tabular}
    \end{center}
    \caption{ \label{Fig:decay2_high_deltav} Normalized decay constant ${\rm Im}(\alpha_2)/k_o$ as a function
    of $\gamma$  when  $\chivt = 19.1^\circ$ and $\delta_v = 16.2^\circ$.  The values of $n_s$ are (1) 1.80, (2) 1.84. }
 \end{figure}

Figures \ref{Fig:decay_SNTF} and \ref{Fig:decay2_high_deltav}) show the decay constants in the anisotropic
partnering material  as  functions of $\gamma$ for for $\tilde{\chi}_v=19.1^\circ$, the remaining parameters 
being the same as in Figure \ref{Fig:v_chiv_19p1}.  The values of $\decayone$ in
Figure \ref{Fig:decay_SNTF}a are larger in the SNTFs than they are in the CTFs for the same values of $n_s$ and $\gamma$.  The values of $\decayone$ appear to decrease towards zero at the upper range of $\gamma$ for each value of $n_s$, except for those cases where the range of $\gamma$ extends to $90^\circ$.  In the two cases where the $\gamma$-range extends to $90^\circ$, $\decayone$ shows a minimum at $90^\circ$.  It should be kept in mind that neither $0^\circ$ nor $90^\circ$ truly represent the limits of a $\gamma$-range  for 
surface--wave propagation, since curves which seem to end at these points are joined to curves of one of the other three, in general, separate $\gamma$-ranges, thereby resulting in only two distinct $\gamma$-ranges.

Values of $\decaytwo$ in the SNTF when $\delta_v=7.2^\circ$ are nearly identical to those in the CTF at a given value of $\gamma$, as shown in Figure \ref{Fig:decay_SNTF}b, for $\tilde{\chi}_v=19.1^\circ$.  The curves of $\decaytwo$ versus $\gamma$ for the various values of $n_s$ nearly join to form a single smooth continuous curve, except near $\gamma=0^\circ$ where the curves for $n_s=1.73$, 1.77 and 1.82 bifurcate.  The value of $\decaytwo$ increases as $n_s$ and $\gamma$ increase.  

The behaviour of $\decaytwo$ for higher modulation amplitude, $\delta_v=16.2^\circ$, is presented in Figure \ref{Fig:decay2_high_deltav} for $n_s=1.80$ and 1.84.  At low values of $\gamma$, $\decaytwo$ is larger for $\delta_v=16.2^\circ$ than for $\delta_v=7.2^\circ$.  As $\gamma$ approaches $90^\circ$, however, the value of $\decaytwo$ for $\delta_v=16.2^\circ$, for both  $n_s=1.80$ and $1.84$, approaches a value only slightly less that observed when $\delta_v=7.2^\circ$ and $n_s=1.96$.

\begin{figure}
    \begin{center}
    \begin{tabular}{c}
    \includegraphics[height=7.3cm]{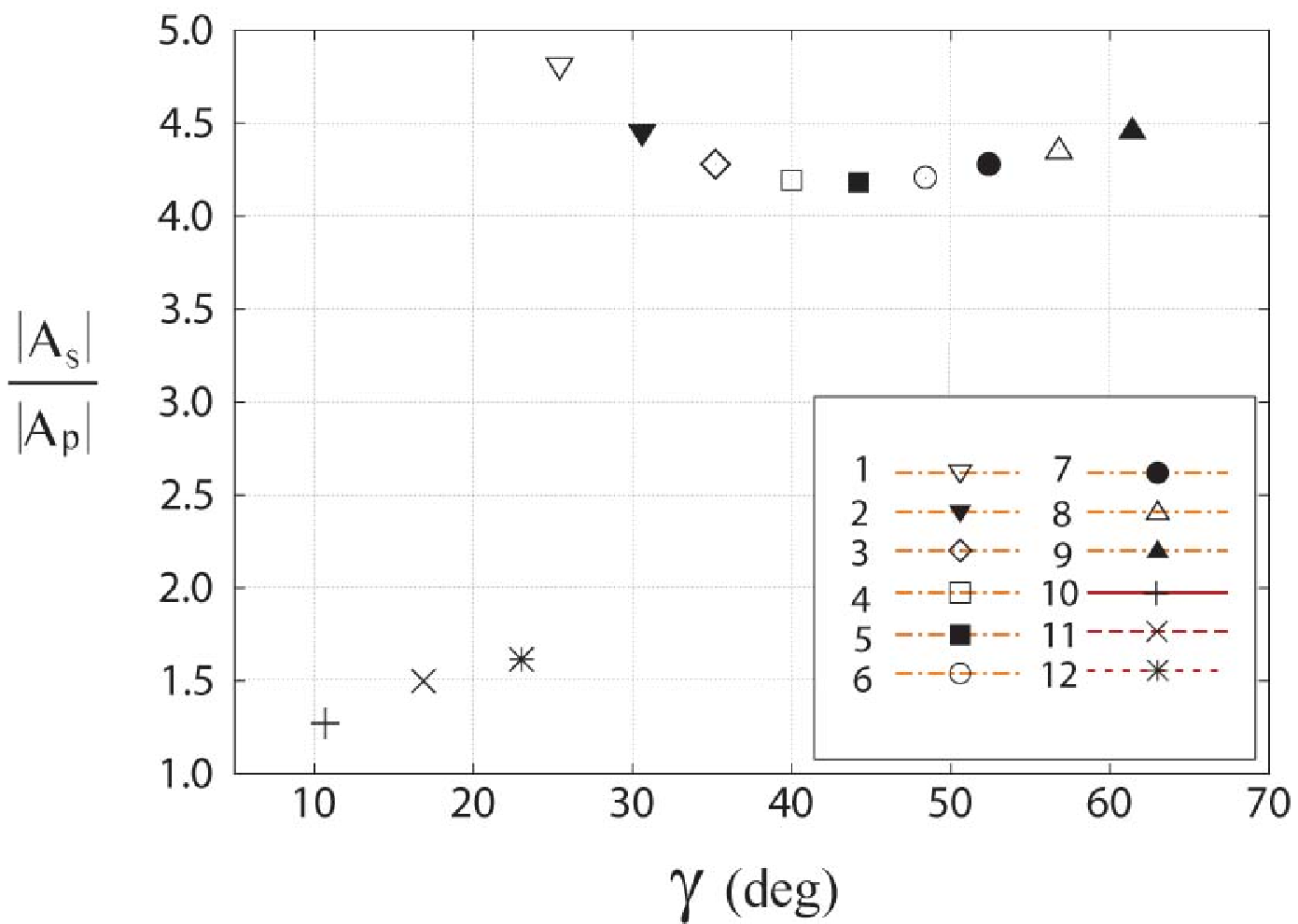}
    \end{tabular}
    \end{center}
    \caption{ \label{Fig:CTF_polarization} Ratio $\vert A_s\vert/\vert A_p \vert$ as a function
    of $\gamma$  when  $\delta_v = 0^\circ$. The values of ${\tilde\chi}_v$ are as follows: (1)-(9) $7.2^\circ$, (10)-(12) $19.1^\circ$. The values of $n_s$ are as follows: (1) 1.57, (2) 1.59, (3) 1.61, (4) 1.63, (5) 1.65, (6) 1.67, (7) 1.69, (8) 1.71, (9) 1.73, (10) 1.80, (11) 1.82, (12) 1.84. }
 \end{figure}

The polarization state of the surface wave in the isotropic partnering material was also investigated. 
First, let us present data when the anisotropic partnering material is a
CTF ($\delta_v=0^\circ$). Figure \ref{Fig:CTF_polarization} shows the ratio of the $s$-polarization amplitude to the the $p$-polarization amplitude  for $n_s=1.57$, 1.59, 1.61, 1.63, 1.65, 1.67, 1.69, 1.71, and 1.73 when $\tilde{\chi}_v=7.2^\circ$; and for $n_s=1.80$, 1.82 and 1.84 when $\tilde{\chi}_v=19.1^\circ$.  When the vapor
incidence angle is low ($7.2^\circ\,\forall z\geq 0$), the surface wave in
the isotropic partnering material is predominantly $s$-polarized with $4<\vert A_s\vert/\vert A_p \vert<5$.  When the vapor
incidence angle is increased to $19.1^\circ\,\forall z\geq 0$, the surface wave is only mildly polarized with $1<\vert A_s\vert/\vert A_p \vert<2$, and $\vert A_s\vert/\vert A_p \vert$ shows a positive slope as a function of $\gamma$.

\begin{figure}
    \begin{center}
    \begin{tabular}{c}
    \includegraphics[height=7.3cm]{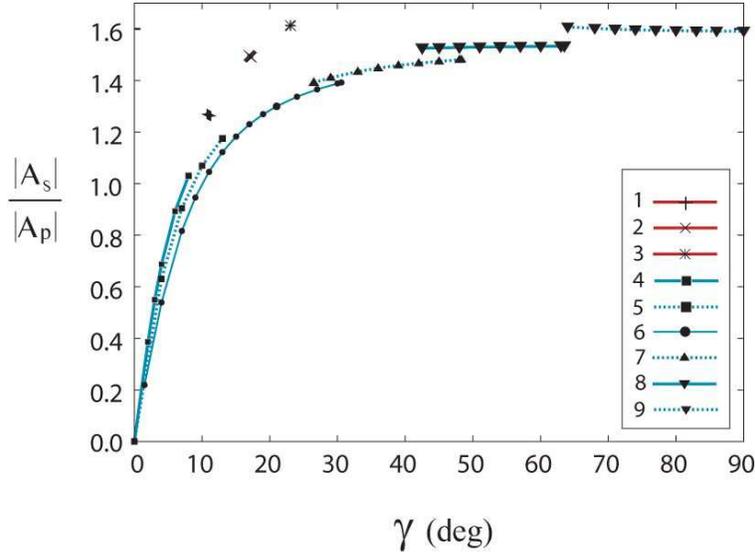}
    \end{tabular}
    \end{center}
    \caption{ \label{Fig:SNTF_polarization} Ratio $\vert A_s\vert/\vert A_p \vert$ as a function
    of $\gamma$  when ${\tilde\chi}_v= 19.1^\circ$. The values of $\delta_v$ are as follows: (1)-(3) $0^\circ$, (4)-(9) $7.2^\circ$. The values of $n_s$ are as follows:  (1) 1.80, (2) 1.82, (3) 1.84, (4) 1.73, (5) 1.77, (6) 1.82, (7) 1.88, (8) 1.92, (9) 1.96. }
 \end{figure}

\begin{figure}
    \begin{center}
    \begin{tabular}{c}
    \includegraphics[height=7.3cm]{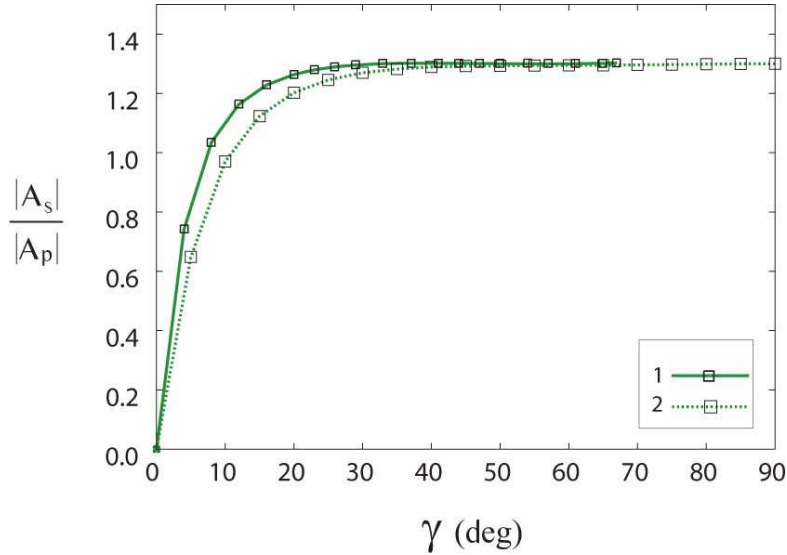}
    \end{tabular}
    \end{center}
    \caption{ \label{Fig:SNTF_polarization_high_deltav} Ratio $\vert A_s\vert/\vert A_p \vert$ as a function
    of $\gamma$  when  ${\tilde\chi}_v = 19.1^\circ$ and $\delta_v = 16.2^\circ$. The values of $n_s$ are (1) 1.80
    and (2) 1.84. }
 \end{figure}

Figure \ref{Fig:SNTF_polarization} shows the ratio $\vert A_s\vert/\vert A_p \vert$ versus $\gamma$, when $\chivt=19.1^\circ$, for CTFs ($n_s=1.80$, 1.82, 1.84) and SNTFs with $\delta_v=7.2^\circ$ ($n_s=1.73$, 1.77, 1.82, 1.88, 1.92, 1.96).  At the same values of $\gamma$, the values of $\vert A_s\vert/\vert A_p \vert$ are slightly smaller for the SNTF than the CTF.  As with the curves describing $\decaytwo$ in Figure \ref{Fig:decay_SNTF}b, the curves describing $\vert A_s\vert/\vert A_p \vert$ for the SNTFs nearly join to form a single, continuous curve.  At $\gamma=0^\circ$, the 
surface wave is entirely $p$-polarized in the isotropic partnering material.  The $s$-polarization 
state intensifies rapidly but then levels off at larger values of $\gamma$.  The value of $\vert A_s\vert/\vert A_p \vert$ is nearly constant at $\sim 1.6$ for $\gamma>60^\circ$.  For $\delta_v=19.1^\circ$, the behaviour of $\vert A_s\vert/\vert A_p \vert$ is similar, as shown in Figure \ref{Fig:SNTF_polarization_high_deltav}, but plateaus at a value of about 1.3 at large values of $\gamma$.

\section{Concluding Remarks}\label{cr}

To conclude, we examined the phenomenon of surface--wave propagation at the planar interface of an isotropic dielectric material and a sculptured nematic thin film with periodic nonhomogeneity.  The boundary--value problem was formulated by marrying the usual formalism for the Dyakonov wave at the planar interface of an isotropic dielectric material and a columnar thin film with the methodology for Tamm states in solid--state physics. The solution of the boundary--value problem led us to predict the existence of Dyakonov--Tamm waves.

Dyakonov surface waves \emph{may} propagate guided by the bimaterial interface of an isotropic dielectric material
and a columnar thin film (or any other biaxial dielectric material). The angular domain of their existence
is very narrow, of the order of a degree. Although several techniques have been suggested \cite{Takayama,NPL-motl},
the widening of that domain has not been impressive.
By periodically distorting the CTF in two ways, either as a chiral STF \cite{LP-jeosrp} or now as a periodically 
nonhomogeneous SNTF, the angular existence domain for surface (Dyakonov--Tamm) waves
can be widened dramatically, even to the maximum possible. A search for Dyakonov--Tamm waves is, at the present time, the most promising route to take for experimental verification of surface-wave propagation at the interface of two dielectric materials at least one of which is  anisotropic.

In Section~\ref{intro}, we posed the following question: Is the huge angular existence 
domain of DyakonovÐTamm waves for the case investigated by Lakhtakia and Polo \cite{LP-jeosrp}
due to the periodicity or due to the structural chirality of the chiral STF? Although the two
attributes are inseparable in a chiral STF, our results in Section~\ref{res} indicate that the
periodic nonhomogeneity is the responsible attribute.

\vspace{0.5 cm}
\noindent {\bf Acknowledgment.} This work was supported in part by the
Charles Godfrey Binder Endowment at Penn State. KA thanks the Department of
Engineering Science and Mechanics, Penn State, and  AL is grateful to the Department
of Physics,  IIT Kanpur, for hospitality.

\vspace{1 cm}

\noindent{\bf References}\\


\begin{thebibliography}{99}
\bibitem{Dyakonov1988}
D'yakonov M I 1988 \textit{Sov. Phys. JETP} \textbf{67} 714

\bibitem{Takayama}
Takayama O, Crasovan L-C, Johansen S K, Mihalache D,
Artigas D and Torner Ll 2008 \textit{Electromagnetics} \textbf{28}, 126

\bibitem{Crasovan}
Crasovan L C, Artigas D, Mihalache D and Torner Ll 2005
\textit{Opt. Lett.} \textbf{30} 3075

\bibitem{Artigas2005}
Artigas D and Torner Ll 2005 \textit{Phys. Rev. Lett.} \textbf{94} 013901


\bibitem{PNL2007}
Polo J A Jr, Nelatury S R and Lakhtakia A 2007 \textit{J. Nanophoton.}
\textbf{1} 013501


\bibitem{AZL1}
Abdulhalim I, Zourob M and  Lakhtakia A 2007 In: 
Marks R, Cullen D, Karube I, Lowe C R and Weetall H H (eds)
\textit{Handbook of Biosensors and Biochips} (Chicester, United Kingdom: Wiley) pp.~413--446

\bibitem{NPL-motl}
Nelatury S R, Polo J A Jr and Lakhtakia A 2008
\textit{Microw. Opt. Technol. Lett.} \textbf{50} 2360

\bibitem{LP-jeosrp}
Lakhtakia A and Polo J A Jr 2007
\textit{J. Eur. Opt. Soc.--Rapid Pub.} \textbf{2} 07021

\bibitem{LMbook}
Lakhtakia A and Messier R 2005
\textit{Sculptured Thin Films: Nano\-engineered Morphology and
Optics} (Bellingham, WA, USA: SPIE Press).

\bibitem{Tamm}
Tamm I 1932 \textit{Phys. Z. Sowjetunion} \textbf{1} 733

 \bibitem{Ohno1990}
Ohno H, Mendez E E, Brum J A, Hong J M, Agull\'o-Rueda F,
Chang L L and Esaki L 1990 \textit{Phys. Rev. Lett.} \textbf{64} 2555

\bibitem{NOA}
Lakhtakia A (ed) 1990
{\em Selected Papers on Natural Optical Acitivity}  
(Bellingham, WA, USA: SPIE)

\bibitem{Charney}
Charney E 1985
{\em The Molecular Basis of Optical Activity} (Malabar, FL, USA: Krieger)

\bibitem{vdH}
van de Hulst H C 1981
{\em Ligh Scattering by Small Particles} (New York, NY, USA: Dover) Sec. 6.4.

\bibitem{TGM}
Mackay T G 2008
 \textit{J. Nanophoton.}
\textbf{2} 029503


\bibitem{Bose}
Bose J C 1898 {\em Proc. R. Soc. Lond. A} {\bf 63} 146

\bibitem{Ro}
Ro R 1991 Determination of the electromagnetic properties of
chiral composites, using normal incidence measurements, Ph.D. dissertation,
Pennsylvania State University, University Park, PA, USA

\bibitem{Bel}
Lakhtakia A 1994
{\em Beltrami Fields in Chiral Media} (Singapore: World Scientific)

\bibitem{Alvaro}
G\'omez \'{A}, Lakhtakia A, Margineda J,  Molina-Cuberos G J,
N\'u$\tilde{n}$ez  J, Ipi$\tilde{n}$a J S, Vegas A and Solano M A 2008 
\textit{IEEE Trans. Microw. Theory Tech.} {\bf 56} 2815

\bibitem{Whites}
Whites K W and   Chung C Y 1997
{\em J. Electromagn. Waves Applics.} {\bf 11} 371

\bibitem{Chan}
Chandrasekhar S 1992
{\em Liquid Crystals} (Cambridge: Cambridge University Press)

\bibitem{HWbook}
 Hodgkinson I J and   Wu Q-h  1997
{\em Birefringent Thin Films and
Polarizing Elements} (Singapore: World Scientific)

\bibitem{Mattox}
Mattox D M 2003
{\em The Foundations of Vacuum Coating Technology} (Norwich, NY, USA:
Noyes Publications)


\bibitem{HWH1998}
Hodgkinson I J, Wu Q H and Hazel J 1998
\textit{Appl. Opt.} \textbf{37} 2653


\bibitem{Chen}
Chen H C 1992
\textit{Theory of Electromagnetic Waves} (Fairfax, VA, USA: TechBooks).

\bibitem{ML2008}
Motyka M A and Lakhtakia A 2008
\textit{J. Nanophoton.} {\bf 2} 021910

\bibitem{YS75}
Yakubovich V A and Starzhinskii V M 1975 \textit{Linear Differential Equations
with Periodic Coefficients} (New York, NY, USA: Wiley).

\bibitem{Jaluria}
Jaluria Y 1996
\textit{Computer Methods for Engineering} (New York, NY, USA: Brunner--Routledge)


\end{thebibliography}
\end{document}